# Unveiling Non-Hermitian Spectral Topology in Hyperbolic Lattices with Non-Abelian Translation Symmetry

Mengying Hu*, Jing Lin, and Kun Ding†

*Department of Physics, State Key Laboratory of Surface Physics, and Key Laboratory of Micro and Nano Photonic Structures (Ministry of Education), Fudan University, Shanghai 200438, China*

*Email: myhu@fudan.edu.cn, †E-mail: kunding@fudan.edu.cn

**Abstract**

The hyperbolic lattice (HBL) has emerged as a compelling platform for exploring matter in non-Euclidean space. Among its notable features, the breakdown of the conventional Bloch theorem stands out, prompting a reexamination of band theory, with the determination of spectra for non-Hermitian systems being a prominent example. Here, we develop an approach to determining the spectra under open boundary conditions (OBCs), one of the foundations in non-Hermitian lattices, from the reciprocal space of HBLs. By introducing supercells to encompass states that are allowed by non-Abelian translational groups, we perform analytic continuation and base on the point gap topology to acquire uniform spectra, the universal OBC spectral range. Applying this method to a single-band nonreciprocal model and a reciprocal non-Abelian semimetal model, we reveal higher-dimensional skin effects and topological phase transitions, respectively, demonstrating the feasibility of our method in predicting spectral topology and investigating non-Hermitian physics in HBLs.

*Introduction*.—Recent years have witnessed a booming development of non-Hermitian physics[1-6], with myriad applications spanning classical waves[7-21], ultracold atoms[22-24], open quantum systems[25-27], and more[28,29]. The breakdown of Hermiticity introduces complex eigenenergies, resulting in spectral topology[5]. Unique to the non-Hermitian system, point gaps (PGs), characterized by eigenvalue winding numbers (EWNs), spontaneously emerge[2,30-35], while Hermitian systems only allow line gaps (LGs). Nontrivial PGs underpin the non-Hermitian skin effect (NHSE)[2,18,34-43], where bulk eigenstates localize at boundaries under open boundary conditions (OBCs) instead of extending as in Hermitian systems. Viewing NHSE from spectra indicates that the eigenenergies under periodic boundary conditions (PBCs) constitute PGs and vary remarkably under OBCs[34,35]. This discrepancy begs for the thermodynamic limit (TDL) OBC spectra calculation at the unit-cell level, breeding the celebrated non-Bloch band theory[2,44-47]. Although the spectral density of states (DOS) is sensitive to details, such as geometry and impurities[43,48-51], the universal OBC spectral range in arbitrary dimensions, denoted as $\sigma_U$, is crucial as it is rooted in the inherent PG topology [52,53]. Both the amoeba theory[52] and uniform spectra formulation[53] confirm $\sigma_U$, underscoring the central role of PGs and analytic continuation of Bloch wavenumbers in non-Hermitian physics.

However, the PG topology is fundamentally broader than that of Bloch wavenumbers, as PGs require only the existence of spectra, whereas Bloch wavenumbers reply on the translational group having one-dimensional (1D) irreducible representations (IRs)—a condition not always satisfied. Hyperbolic lattices (HBLs), constructed on a two-dimensional hyperbolic plane with constant negative curvature, exemplify such cases[54], differing from Euclidean lattices (ELs). Recent experimental achievements have positioned HBLs as a brilliant platform to investigate quantum matter in non-Euclidean geometries[55,56]. Hyperbolic counterparts of phenomena like topological insulators[57-63], Hofstadter states[57,64,65], flat bands[66-68], strong-correlated states[69,70], and Anderson localization[71,72] have been proposed, with some realized experimentally[58,60,61,63,70]. Exploring physics and phenomena unique to HBLs or lacking Euclidean analogs remains particularly compelling.

Unlike ELs, HBLs stand out due to negative curvature, and ensuingly, their translational groups, which are cornerstones in analyzing lattices, have been fundamentally altered[73]. Hyperbolic translation groups, known as Fuchsian groups, are non-Abelian and admit higher-dimensional IRs, necessitating a generalization of the Bloch theorem[74-76]. This demand makes hyperbolic band theory (HBT) take shape, which incorporates Abelian [U(1)] and non-Abelian [U($d$) for dimensions $d > 1$] components, corresponding to Abelian and non-Abelian Bloch states (NABSs)[74-80]. Determining the TDL spectra becomes first and foremost because the presence of NABSs invalidates standard EL approaches. Recently, several methods have been successfully established for forecasting Hermitian DOS under TDL, which reveals the criticality of NABSs [77,78]. While exceptional contours[81] and NHSEs[82-84] have been investigated in specific non-Hermitian HBLs, the general determination of uniform spectra $\sigma_U$ with NABSs included remains an open and critical challenge.

To address this, we here formulate a systematic approach to determine $\sigma_U$ for non-Hermitian HBLs. Our method utilizes incremental supercells to incorporate more NABSs and further employs analytic continuation to obtain $\sigma_U$. To illustrate, we first analyze a single-band model to contrast the PBC spectra and $\sigma_U$, thereby demonstrating the NHSE. We then delve into a celebrated non-Abelian semimetal model and reveal that NABSs exhibit distinct non-Hermitian phases compared with Abelian Bloch states (ABSs)[85].

***Determination of uniform spectra***.—To detail the conundrum of acquiring $\sigma_U$ in HBLs, we commence from lattice tessellation to translation groups. Infinite lattices are typically tessellated into unit cells related by translational operations. Beyond using the smallest primitive cell, one can adopt specific groupings of $n$ ($> 1$) primitive cells, dubbed *n-supercell* ($n = 1$ being the primitive cell), to tile the lattice. Figures 1(a) and 1(b) depict the primitive cell, 2-supercell, and translational operators $\gamma_i$ of a Euclidean {4,4} and hyperbolic {8,8} lattice (see details in Sec. I, Ref. [86]). PBCs compactify these cells into closed manifolds [bottom in Fig. 1(a,b)], referred to as PBC clusters. For an EL, any unit cell, regardless of *n*, is compactified to a torus [Fig. 1(a)]. However, for an HBL, the genus $g^{(n)}$ ($= n(g^{(1)} - 1) + 1$) of an *n*-supercell increases linearly with *n*, meaning different supercells are compactified onto manifolds with different genera[87,88]. For instance, the primitive cell and the 2-supercell in Fig. 1(b) are compactified onto surfaces with $g^{(1)} = 2$ and $g^{(2)} = 3$, respectively. This distinction reflects inherent differences in translation groups of HBLs and ELs[76].

Translational operations associated with primitive cells build the maximal translation group $\Gamma^{(1)} = \Gamma$, while those on *n*-supercells form subgroups $\Gamma^{(n)} \subset \Gamma^{(1)}$. The commutativity of $\gamma_i \in \Gamma^{(1)}$ in ELs ensures Abelian translation groups with only 1D IRs, forming the basis of the conventional Bloch theorem[89]. The *n*-supercells merely reduce the Brillouin zone size, resulting in band folding without new physics, making primitive cell calculations sufficient in the TDL. In contrast, $\gamma_i \in \Gamma^{(1)}$ in HBLs generally do not commute, and translation groups allow higher-dimensional IRs[76]. These characterize states that cannot simply be modeled by a Bloch phase under translation, even with analytic continuations, thus defining NABSs[76]. Applying PBCs to the primitive cell only captures ABSs within the U(1)-HBT, which is insufficient for forecasting the TDL spectra. Employing $n$-supercells as PBC clusters enables the identification of NABSs because investigating $\Gamma^{(n)}$ can voluntarily invoke higher-dimensional IRs of $\Gamma$[78]. Hence, constructing a sequence of such PBC clusters to approach the TDL is then feasible and has shown a rapid convergence in the DOS for Hermitian systems as $n$ increases, consistent with real-space calculations[77].

We now turn to the non-Hermitian scenario, where calculating $\sigma_U$ is central. Since the essence of $n$-supercells lies in the necessity of examining $\Gamma^{(n)}$ as $n$ increases, together with the fact that $\sigma_U$ is based on the analytic continuation of Bloch wavenumbers into the complex plane, we next generalize the supercell method[78] to accommodate non-Hermitian HBLs. Wrapping the $n$-supercell into a PBC cluster requires implementing $d^{(n)}_{\{p,q\}}$ ($= 2g^{(n)}$) numbers of $e^{ik_j}$ ($j = 1, \ldots, d^{(n)}_{\{p,q\}}$) to specific pairs of boundaries (see details in Sec. I, Ref. [86]), and we perform $k_j \to k_j - i\mu_j$ for each $k_j$, leading to the $n$-supercell Hamiltonian $H^{(n)}(\beta_j = e^{ik_j+\mu_j})$. The EWNs are defined and now depend on $n$ as follows[33-35]:

$$w_\alpha^{(n)}(E) = \int_0^{2\pi} \frac{dk_\alpha}{2\pi i} \partial_{k_\alpha} \ln f^{(n)}(E, \boldsymbol{\beta}^{(n)}),  \quad (1)$$

where $E \in \mathbb{C}$ is the base energy, $\alpha = 1, \ldots, d^{(n)}_{\{p,q\}}$, $f^{(n)}(E, \boldsymbol{\beta}^{(n)}) = \det[H^{(n)}(\boldsymbol{\beta}^{(n)}) - E]$ is the characteristic polynomial for the $n$-supercell, and $\boldsymbol{\beta}^{(n)} = \{\beta_j; j = 1, \ldots, d^{(n)}_{\{p,q\}}\}$. The dependence of $w_\alpha^{(n)}(E)$ on $k_\alpha$ arises because winding numbers are defined over a parametric loop. Since nonzero EWNs forecast the occurrence of PGs, we introduce the rescaled spectrum of $n$-supercells as[35,53]:

$$\mathrm{Sp}_\ell^{(n)}(|\beta_j|) = \{E; \sum_\alpha |w_\alpha^{(n)}(E)| \neq 0\}, \quad (2)$$

where each $\ell$ corresponds to a specific configuration of $|\beta_j|$ values. Physically, $\mathrm{Sp}_\ell^{(n)}(|\beta_j|)$ identifies the spectral range with nonzero total EWNs for a given set of $|\beta_j|$ values[53]. The energy

residing in $\sigma_U^{(n)}$ requires that the total EWNs do not vanish for all $|\beta_j|$, or equivalently, PGs are nontrivially open [35]. Hence, $\sigma_U^{(n)}$ is obtained by taking the intersection of all $\text{Sp}_\ell^{(n)}(|\beta_j|)$ as

$$\sigma_U^{(n)} = \bigcap_{|\beta_j| \in (0,\infty),j} \text{Sp}_\ell^{(n)}(|\beta_j|). \tag{3}$$

Figure 1(c) sketches this intersection process, where three $\text{Sp}_\ell^{(n)}$ are shown as representatives of all $\text{Sp}_\ell^{(n)}$. The red hatched area, representing the intersection, corresponds to $\sigma_U^{(n)}$. Incrementing $n$ then determine $\sigma_U$ in non-Hermitian HBLs.

To demonstrate this approach, we first deploy the single-band models in the Euclidean {4,4} and hyperbolic {8,8} lattice. Figures 1(d) and 1(e,f) display the spectral ranges obtained from various $n$-supercells for the EL and HBL, respectively. The $n$-supercell PBC spectra $\sigma_P^{(n)}$ (grey areas) are calculated from the PBC cluster approach[78], while $\sigma_U^{(n)}$ (red areas) are obtained by our method. A common feature across Figs. 1(d-f) is $\sigma_U^{(n)} \subset \sigma_P^{(n)}$ for all $n$, a hallmark of NHSEs. As expected, the values of $n$ do not alter $\sigma_P^{(n)}$ and $\sigma_U^{(n)}$ in the ELs [Fig. 1(d)], which aligns with the Abelian nature of $\Gamma^{(1)}$. Comparing Figs. 1(e) and 1(f) shows that $\sigma_P^{(n)}$ and $\sigma_U^{(n)}$ do not vary significantly with $n$, which is a consequence of the symmetry and nonreciprocity used in this concrete model (see details in Sec. II, Ref. [86]). We will later demonstrate the non-Abelian characteristic, where $\sigma_{P/U}^{(1)} \neq \sigma_{P/U}^{(n \neq 1)}$, highlighting the non-Abelian nature of $\Gamma^{(1)}$. Moreover, the observed differences between $\sigma_U^{(1)}$ and $\sigma_U^{(2)}$ illustrate the efficiency of our method (see details in Sec. II, Ref. [86]), further confirming the NHSE feature $\sigma_U^{(n)} \subset \sigma_P^{(n)}$. This result underscores that our method provides a systematic framework for investigating the spectral topology of HBLs, particularly the PG topology and phase transitions, which will be discussed next.

*Higher-dimensional skin effects*.—We first scrutinize the NHSE, a renowned manifestation of PG topology, to validate our recipe. Considering that higher-dimensional IRs possibly blur skin effects, we employ a particular kind of structure with meticulous boundary connections, which can be described using U(1)-HBT. As such, the primitive cell suffices to reflect the essential properties, and we name them as higher-dimensional Euclidean lattices (HDELs)[63]. To illustrate this, we deploy the nonreciprocal single-band model used in Fig. 1 on the HDEL and plot the spectra as blue circles in Fig. 2(a), which clearly reside within $\sigma_P^{(1)}$. By substituting quantized 4-dimensional momenta into $H^{(1)}(k_j)$, we compute the discretized PBC spectrum using U(1)-HBT, which aligns perfectly with the blue circles in Fig. 2(a) (see details in Sec. II, Ref. [86]). This agreement confirms that $\sigma_\text{P}^{(1)}$ is sufficient for modeling HDELs, suggesting that the HDEL serves as an alternative platform for exploring non-Hermitian phenomena in high-dimensional Euclidean systems. The white and green lines in Fig. 2(a) represent the PBC spectra of the primitive cell by varying $k_1$ while fixing other $k_j$. Nontrivial PGs are clearly seen, and the occurrence of NHSE is forecasted.

By removing the PBC connections in the HDEL, we have the corresponding OBC structure, and the blue circles in Fig. 2(b) depict the calculated OBC spectra. For comparison, $\sigma_\text{U}^{(1)}$ from Fig. 1(e), which defines the spectral range where Abelian states can exist, is also shown. The OBC spectra lie within $\sigma_\text{U}^{(1)}$ and are noticeably distinct from the PBC spectra, suggesting NHSE. An exhibition of the nonreciprocal NHSE is the skewness of right and left eigenvectors[90,91]. We thus use the Kullback–Leibler (KL) divergence, a non-negative measure that quantifies the difference between two distributions, equaling 0 when they are identical and increasing with greater disparity[92]. The inset of Fig. 2(b) presents the percentage distribution of KL divergence between the right and left eigenvectors for states within the range highlighted by the white arrow, indicating the presence of NHSE.

Besides features in the eigenstates, NHSEs still highlight themselves as directional amplifications during transport[93-96]. Hence, for showcasing the NHSE, we choose to employ the Green's function $G(E) = (E - H_\text{OBC})^{-1}$, where $H_\text{OBC}$ is the OBC Hamiltonian, and $E$ is an arbitrary energy. The upper panel of Fig. 2(c) illustrates the distribution of $|\langle p|G(E = 2.5)|s\rangle|$ as functions of $s$ and $p$, wherein both are placed along the $\gamma_1$ direction, and the marker sizes are linearly proportional to their magnitudes. A clear tendency toward the negative $\gamma_1$ direction is seen, aligning with the spectral sum of eigenstates within the same range, as shown in the lower panel of Fig. 2(c). This tendency, together with the eigenstates, confirms the occurrence of NHSE, consistent with the nonreciprocal coupling direction. For comparison, Fig. 2(d) exhibits the distribution of $|\langle p|G(E = 2.5)|s\rangle|$ and the spectral sum of eigenstates along the $\gamma_2$ direction. No directional tendency is seen in the Green's function, corroborating the absence of NHSE in the eigenstates. The contrast between the $\gamma_1$ and $\gamma_2$ directions attests to the NHSE and PG topology, verifying the implication of uniform spectra obtained by our method in HBLs.

*Non-Abelian semimetals*.—LGs, though also present in Hermitian systems, are as crucial as PGs for understanding eigenstate behaviors in non-Hermitian systems, as their spectral range closely relates to topological phase transitions in the eigenstates. Notably, it has been revealed in Hermitian HBLs that the emergence of NABSs can be even dominant over ABSs during such transitions[80]. To illustrate this, we employ the celebrated non-Abelian semimetal model on the {8,8} lattice, defined as (see details in Sec. III, Ref. [86]):

$$H_{\text{OBC}} = \sum_{r,s} \left[\psi_r^\dagger \frac{\Gamma_5 - i\Gamma_s}{2}\psi_{r+s} + \text{h.c.}\right] + m\sum_r \psi_r^\dagger \Gamma_5 \psi_r, \tag{4}$$

where the spinor $\psi_r$ has four components at each site $r \in \mathbb{Z}^4$, $s\ (=1,\ldots,4)$ corresponds to four translational directions, $\{\Gamma_\nu\}_{\nu=1}^5 = \{\sigma_1 \otimes \sigma_0, \sigma_2 \otimes \sigma_0, \sigma_3 \otimes \sigma_1, \sigma_3 \otimes \sigma_2, \sigma_3 \otimes \sigma_3\}$ with $\sigma_\nu$ being the Pauli matrices, and $m \in \mathbb{C}\ (= m_r + im_i)$ denotes the onsite potential. When $m \in \mathbb{R}$, this model can be treated as the hyperbolic counterpart of the four-dimensional quantum Hall insulator in ELs, but the presence of NABSs spoils the LGs with nonzero second Chern numbers ($C_2 \neq 0$) from ABSs and makes it become semimetal. Concisely, at $m_r = 4$, the primitive cell (2-supercell) displays a topological phase transition from $C_2 \neq 0$ (semimetal) to a trivial insulator. Determining $\sigma_U$ with NABSs included is now required to investigate how $m_i$ alters the transitions, which are shown by the blue and red squares in Fig. 3(a). The left and right panels respectively depict results for the primitive cell and 2-supercell (see details in Sec. III, Ref. [86]). We can see that both transition points in $m_r$ decrease when $m_i$ climbs up, and thus, it can be expected that for a given $m$, $\sigma_U^{(2)}$ must show distinct behaviors compared with $\sigma_U^{(1)}$, a remarkable feature of NABSs.

We first choose $m = 3 + 0.3i$, which lies in the primitive cell gapped (2-supercell gapless) phase. The calculated $\sigma_U^{(1)}$ ($\sigma_U^{(2)}$) in Fig. 3(b) confirms that the LG is open (close) when non-Abelian states are absent (present). To further verify the origin of LG closing, we call U(2)-HBT[78,97] and show its spectra by grey dots in Fig. 3(b). Good agreement in the gap closing validates our method of $\sigma_U$ and confirms that the spectral behaviors of non-Abelian states are from NABSs for reciprocal non-Hermitian systems (see Sec. III in Ref. [86]). We now turn to the trivial phase for both the primitive cell and the 2-supercell. Setting $m = 5 + 0.5i$, we clearly see a gap in $\sigma_U^{(1)}$ [Fig. 3(c)]. We also depict $\sigma_U^{(4)}$ in Fig. 3(c), which showcases that the LG remains intact in the TDL. However, the non-emptiness of $\sigma_U^{(4)} - \sigma_U^{(1)}$ hints at the appearance of non-Abelian states under OBCs. Thus, we compute the OBC spectra for two configurations, represented by the filled circles and stars in the inset of Fig. 3(c). Both OBC spectra lying in $\sigma_U^{(4)}$ validate our method of acquiring $\sigma_U$, and degenerate states within $\sigma_U^{(4)} - \sigma_U^{(1)}$ affirm the existence of non-Abelian states. The distinction between the primitive cell and the *n*-supercell highlights the critical role of non-Abelian states in forming LGs and determining the OBC spectral range, further emphasizing the criticality of accurately calculating $\sigma_U$.

*Discussions and conclusions*.—In summary, we implement incremental supercells that can be wrapped into PBC clusters to encompass non-Abelian states and perform analytic continuation to determine uniform spectra in HBLs based on the connotation of PGs. By applying it to a single-band nonreciprocal model, we demonstrate higher-dimensional skin effects in an HDEL, showcasing the PG topology. Furthermore, through a non-Abelian semimetal model, our method successfully forecasts the topological phase transition points in both Abelian and non-Abelian states. Therefore, our approach offers a feasible way to investigate spectral topology and non-Hermitian phase transitions associated with gaps in HBLs. While our focus is on acquiring OBC spectra in the sense of TDL, the underlying principles of our method, which are based on subgroups and induced representations, can be generalized to other non-periodic lattices with well-defined generation groups, such as Cayley tree [98], Bethe lattice[99], and fractals[100]. Given recent advances in non-Hermitian topological invariants of higher-dimensional systems, regardless of whether non-Bloch [52] or real-space approaches[101,102] are employed, we believe our method is a valuable tool for investigating topological phase transitions and quantum phenomena in non-Hermitian HBLs.

**Acknowledgment**

We thank Prof. C. T. Chan and Mr. Nan Cheng for the helpful discussions. This work is supported by the National Key R&D Program of China (No. 2022YFA1404500, No. 2022YFA1404701), the National Natural Science Foundation of China (No. 12174072, No. 2021hwyq05), and the China Postdoctoral Science Foundation (No. 2023M730705).

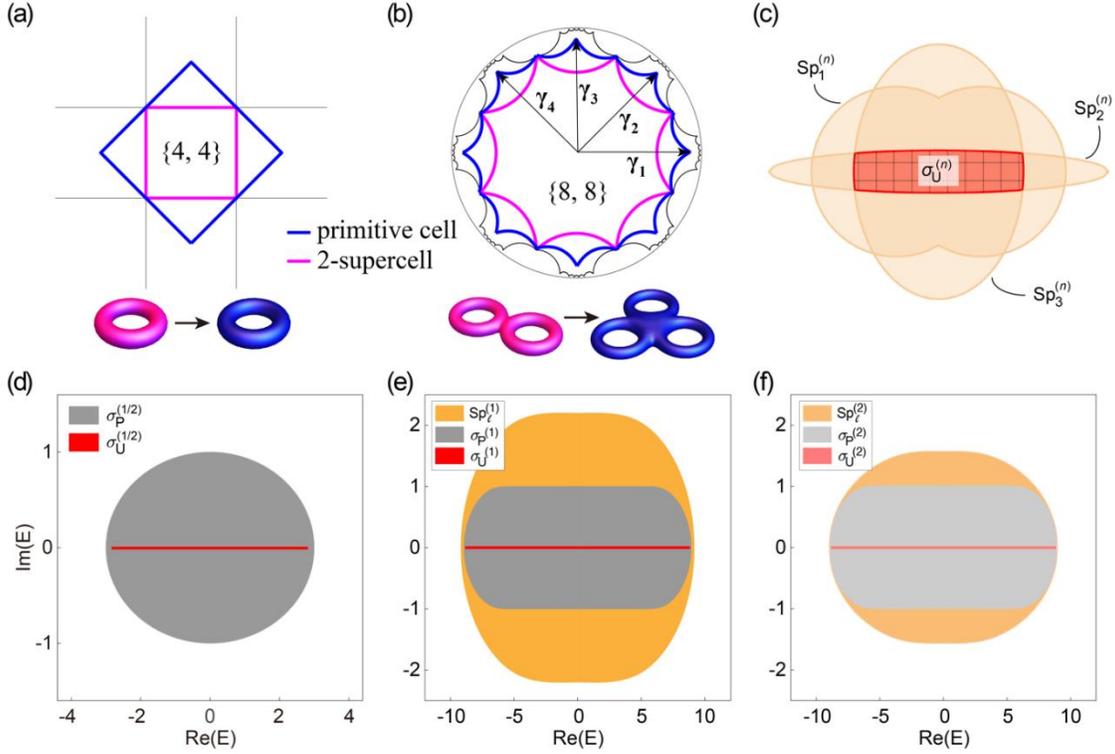

FIG. 1. (a,b) Symmetrized construction of *n*-supercells in (a) the Euclidean {4,4} and (b) hyperbolic {8,8} lattice. Translational operators $\{\gamma_i\}$ for each lattice are labeled herein. The corresponding translation groups are $\Gamma^{(1)}_{\{4,4\}} = \langle \gamma_1, \gamma_2 \mid \gamma_1 \gamma_2^{-1} \gamma_1^{-1} \gamma_2 = 1 \rangle$ and $\Gamma^{(1)}_{\{8,8\}} = \langle \gamma_1, \gamma_2, \gamma_3, \gamma_4 \mid \gamma_1 \gamma_2^{-1} \gamma_3 \gamma_4^{-1} \gamma_1^{-1} \gamma_2 \gamma_3^{-1} \gamma_4 = 1 \rangle$. The region enclosed by the magenta (blue) line at the top is the primitive cell (2-supercell), with its corresponding compactification sketched at the bottom. (c) Schematic illustration of our approach to calculate $\sigma_U^{(n)}$. $\mathrm{Sp}_1^{(n)}$, $\mathrm{Sp}_2^{(n)}$, and $\mathrm{Sp}_3^{(n)}$ are depicted to represent all $\mathrm{Sp}_\ell^{(n)}$, with their intersection (red hatched area) corresponding to $\sigma_U^{(n)}$. (d-f) $\sigma_P^{(n)}$ (grey areas) and $\sigma_U^{(n)}$ (red areas) for different supercells in (c) the {4,4} and (d) {8,8} lattice. The dark and light colors stand for the $n=1$ and $n=2$ cases, respectively. The orange areas in (e) and (f) display typical patterns of $\mathrm{Sp}_\ell^{(n)}$ for $|\beta_j| = \{0.7, 0.7, 0.7, 0.7\}$ and $|\beta_j| = \{1.1, 1.2, 0.9, 1.2, 1.1, 0.8\}$, respectively. The EL model used is $H(\beta_1, \beta_2) = \beta_1 + \frac{1}{2}\beta_1^{-1} + \beta_2 + \frac{1}{2}\beta_2^{-1}$, while the HBL model is $H(\beta_1, \beta_2, \beta_3, \beta_4) = 2\beta_1 + \beta_1^{-1} + \beta_2 + \beta_2^{-1} + \beta_3 + \beta_3^{-1} + \beta_4 + \beta_4^{-1}$.

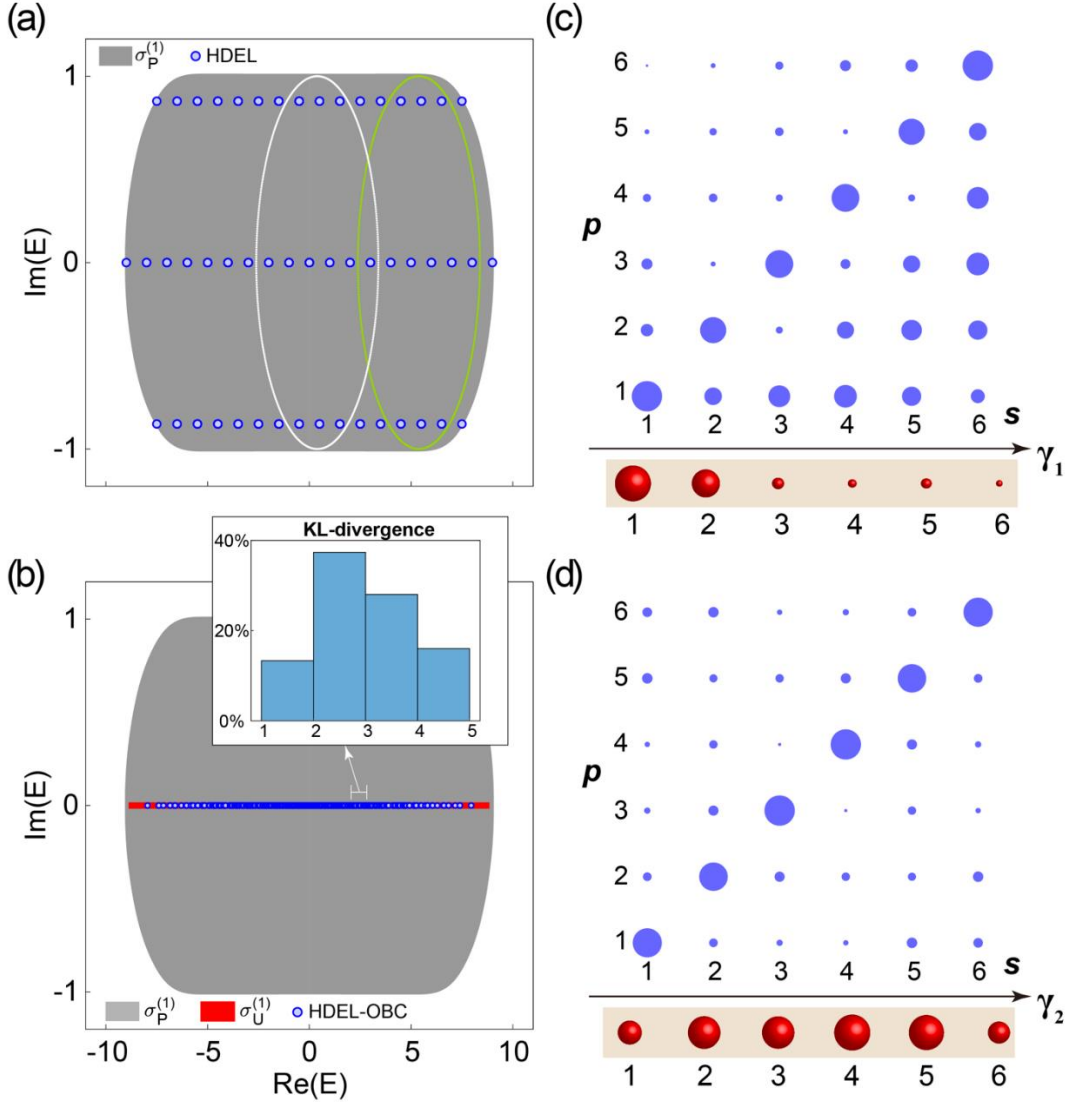

FIG. 2. (a,b) Spectra (blue circles) of (a) an HDEL and (b) the corresponding OBC configuration. The regions $\sigma_P^{(1)}$ and $\sigma_U^{(1)}$ are plotted in grey and red, respectively. The values of $\{k_2, k_3, k_4\}$ used in the white and green lines of (a) are $\{\pi/7, \pi/7, \pi/7\}$ and $\{\pi/4, 2\pi/3, \pi/2\}$, respectively. (c,d) Distribution of Green's functions (top panels) and the spectral sum of eigenstates (bottom panels) along (c) the $\gamma_1$ and (d) the $\gamma_2$ directions, with the marker sizes proportional to their magnitudes. The excitation energy is $E = 2.5$. The HDEL contains 1296 sites with explicit geometry defined in Sec. II of Ref. [86], and other parameters are the same as in Fig. 1.

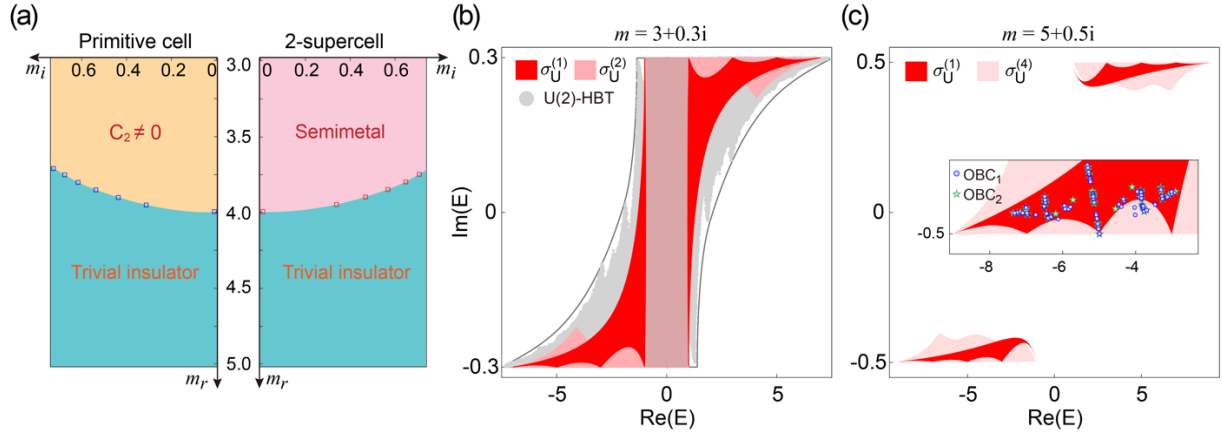

FIG. 3. (a) Phase diagram of the primitive cell (left) and 2-supercell (right) in the complex *m* plane. The regions in light orange and pink highlight the $C_2 \neq 0$ phase and semimetal phase, while the light cyan regions denote the trivial insulator phase. Blue and red squares represent the calculated phase transition points for the primitive cell and 2-supercell, respectively. (b) $\sigma_U^{(1)}$ (red) and $\sigma_U^{(2)}$ (light red) when $m = 3 + 0.3i$. The filled circles represent the spectra from U(2)-HBT by gathering 30 ensembles, and the dark solid line highlights their spectral boundary. (c) $\sigma_U^{(1)}$ (red) and $\sigma_U^{(4)}$ (light red) when $m = 5 + 0.5i$. The inset shows the lower branch of the spectra with the blue and green markers depicting the OBC spectra for two configurations consisting of 1601 and 1142 sites.